\documentclass[12pt]{article}
\usepackage{a4}
\usepackage{amsbsy}
\usepackage{epsfig}

\newlength{\abstwidth}
\newcommand{\RPP}[3]{{\it Rept.\ Prog.\ Phys.\ }{\bf #1} (#2) #3}

\newcommand{\PRD}[3]{{\it Phys.\ Rev.\ }{\bf D#1} (#2) #3}
\newcommand{\PLB}[3]{{\it Phys.\ Lett.\ }{\bf B#1} (#2) #3}

\newcommand{\NPB}[3]{{\it Nucl.\ Phys.\ }{\bf B#1} (#2) #3}
\newcommand{\ZPC}[3]{{\it Z.\ Phys.\ }{\bf C#1} (#2) #3}

\newcommand{\hep}[1]{{\rm hep-ph/#1}}
\newcommand{\hex}[1]{{\rm hep-ph/#1}}

\newcommand{\pT}{p_{{\mathrm T}}}
\newcommand{\kT}{k_{{\mathrm T}}}

\newcommand{\vqoT}{{\vec{q}_{1{\mathrm T}}}}
\newcommand{\vqtT}{{\vec{q}_{2{\mathrm T}}}}
\newcommand{\qoT}{q_{1{\mathrm T}}}
\newcommand{\qtT}{q_{2{\mathrm T}}}

\renewcommand{\d}{{\mathrm d}}

\newcommand{\e}{{\mathrm e}}
\newcommand{\f}{{\mathrm f}}

\newcommand{\p}{{\mathrm p}}
\newcommand{\q}{{\mathrm q}}

\newcommand{\qbar}{\bar{\mathrm q}}

\newcommand{\F}{{\mathrm F}}

\newcommand{\T}{{\mathrm T}}
\renewcommand{\L}{{\mathrm L}}

\newcommand{\bJ}{{\boldsymbol{J}}}
\newcommand{\bP}{{\boldsymbol{P}}}
\newcommand{\beq}{{\boldsymbol{=}}}
\newcommand{\bzero}{{\boldsymbol{0}}}
\newcommand{\bone}{{\boldsymbol{1}}}
\newcommand{\btwo}{{\boldsymbol{2}}}

\newcommand{\bminus}{{\boldsymbol{-}}}
\newcommand{\bplus}{{\boldsymbol{+}}}

\begin{document}

\pagestyle{empty}

\begin{flushright}
CERN--TH/99--131\\ 
hep-ph/9905305
\end{flushright}

\vspace{\fill}

\begin{center}
{\Large\bf 
Evidence that the Pomeron transforms \\[1ex]
as a non-conserved vector current}\\[1.8ex]
{\bf Frank E.\ Close${}^{a}$} 
\\[1.0ex]
and \\ [1.5ex]
{\bf Gerhard A.\ Schuler${}^{b}$} \\[2.0ex] 
Theoretical Physics Division, CERN, CH-1211 Geneva 23 \\
\end{center}

\vspace{\fill}
\begin{center}
{\bf Abstract}\\[2ex]
\begin{minipage}{\abstwidth}
The detailed dependences of central meson production on
the azimuthal angle $\phi$, $t$ and the meson $J^P$ 
are shown to be consistent with the hypothesis that the soft Pomeron
transforms as a non-conserved vector current. Further tests 
are proposed. This opens the way for a quantitative 
description of $\q\qbar$ and glueball production in 
$\p\p \to \p\, M\, \p$. 
\end{minipage}
\end{center}

\vspace{\fill}
\noindent
CERN--TH/99--131\\
May 1999

\vspace{\fill}
\noindent
\rule{60mm}{0.4mm}

\vspace{0.1mm} 
\noindent
${}^a$ On leave from Rutherford Appleton Laboratory, 
Chilton, Didcot, Oxfordshire, OX11 0QX, UK;
supported in part by 
the EEC-TMR Programme, Contract NCT98-0169.
\\[1.5ex]
${}^b$ Heisenberg Fellow; 
supported in part by the EU Fourth Framework Programme
``Training and Mobility of Researchers'', Network 
``Quantum Chromodynamics and the Deep Structure of Elementary Particles'', 
contract FMRX-CT98-0194 (DG 12-MIHT). 
\clearpage
\pagestyle{plain}
\setcounter{page}{1} 


\section{Introduction}
In order to understand the dynamics of the proposed glueball 
filter\cite{CK} and to separate glueballs from $\q\qbar$ states in 
central production \cite{Robson} it is necessary to establish the 
transformation properties of the Pomeron at low momentum transfers. 
The observations \cite{WA} of non-trivial dependence on the
azimuthal angle $\phi$ of the outgoing protons in 
$\p\p \to \p\p + M$
shows that the effective spin of the Pomeron cannot be 
simply zero \cite{Us,Diehl}.
Recently we have shown \cite{Us} that many features of
the central production of several established $\q\qbar$ mesons 
in $\p\p \to \p\p + M$, in both the azimuthal, $\phi$, 
and glueball filter, $\kT$, dependence 
can be understood if the Pomeron behaves as a conserved vector current. 
However, there are both empirical \cite{Us} and theoretical 
reasons \cite{Diehl,kaid} to believe that this cannot be the whole story 
and that current {\bf non}-conservation is important, 
especially at the central meson production-vertex. 

In this letter we show that the data clearly require that the Pomeron 
transforms effectively as a non-conserved vector with a behaviour 
of a specific type. 
We propose further tests of this hypothesis and discuss the 
practicalities of differentiating glueballs from $\q\qbar$ in 
$0^{++},2^{++}$ meson production.

As in our study of the conserved vector-current 
(CVC) case \cite{Us} we consider 
the central production of a $J^{P+}$ meson $M$ in 
high-energy proton--proton scattering. We have shown, 
with current conservation at the 
proton--Pomeron vertex (see also \cite{Diehl}),
that the cross section may be written as
\begin{equation}
  \d\sigma/ \d t_1\, \d t_2\, \d\phi\, \d x_{\F}
       \sim t_1\, t_2\, 
           \left( \sigma_2 +  \sigma_1 +  \sigma_0 \right)
\ ,
\label{eq:dsigde}
\end{equation}
where 
the subscripts $i$ denote the helicity states of the meson and 
the suppressed pre-factor has the following properties: 
it is concentrated at 
$x_{\F} \approx 0$, 
steeply falling with decreasing four-momentum transfers, $t_i$, 
and finite for $t_i$ approaching $t_i^{\min} \approx 0$
(i.e.\ approximately proportional to $\exp\{b(t_1 + t_2)\}$). 

In the kinematic regime of interest, $|t_i| \ll M^2$ and 
$x_i \approx (\sqrt{x_{\F}^2 + 4 M^2/s} \pm x_{\F})/2 \ll 1$, 
we have
\begin{eqnarray}
  \sigma_2 & = &   \frac{1}{2}\, A_{+-}^2
\nonumber\\ 
  \sigma_1 & = &   A_{+\L}^2   + A_{\L+}^2
    - 2\, \eta\,  \xi_2\, A_{+\L}\, A_{\L+} \, \cos \phi
\nonumber\\ 
  \sigma_0 &  = &  
    \left( A_{\L\L} - \xi_1 \,  A_{++}\, \cos \phi \right)^2
\qquad (\eta = +1)
\nonumber\\ & = & A_{++}^2\, \sin^2 \phi
\qquad\qquad\qquad (\eta = -1)
\ ,
\label{eq:approxsig}
\end{eqnarray}
where the subscripts $\pm$ and $\L$ refer to the Pomeron helicities, 
$\xi_i$ are sign factors, and $\eta$ is the product of the
naturality of the meson and the two currents, 
$\eta = \eta_1\, \eta_2\, \eta_M = \pm 1$.
The general structure of the $\phi$ dependence as a function of $J^P$ is
then as follows, from which we will abstract specific tests:
\begin{eqnarray}
 \d\sigma[0^-] & \sim & t_1\, t_2\, A_{++}^2\, \sin^2 \phi
\label{eq:zminus}
\\
 \d\sigma[0^+] & \sim & \left( \sqrt{t_1\, t_2}\, A_{\L\L} 
  - \xi_1\, \sqrt{t_1\, t_2}\, A_{++} \, \cos\phi \right)^2
\label{eq:zplus}
\\
 \d\sigma[1^+] & \sim & t_1\, t_2\, A_{++}^2\, \sin^2 \phi
 + \left( \alpha\, \vqoT - \beta\, \vqtT \right)^2
\nonumber\\  & & \hspace*{-3ex}
(\alpha = \qtT\, A_{+ \L}, \beta = \eta\xi_2\, \qoT\, A_{\L +}
\ \mathrm{or}\ \beta = \qoT\, A_{+ \L}, \alpha = \eta\xi_2\, \qtT\, A_{\L +} )
\label{eq:oplus}
\\
 \d\sigma[2^+] & \sim & \left( \sqrt{t_1\, t_2}\, A_{\L\L} 
  - \xi_1\, \sqrt{t_1\, t_2}\, A_{++} \, \cos\phi \right)^2
\nonumber\\
 & &~ 
 + \left( \alpha\, \vqoT - \beta\, \vqtT \right)^2 
 + t_1\, t_2\, \frac{1}{2}\, A_{+-}^2
\label{eq:tplus}
\\
 \d\sigma[2^-] & \sim &  t_1\, t_2\, A_{++}^2\, \sin^2 \phi
\nonumber\\
 & &~ 
 + \left( \alpha\, \vqoT - \beta\, \vqtT \right)^2 
 + t_1\, t_2\, \frac{1}{2}\, A_{+-}^2
\ .
\label{eq:tminus}
\end{eqnarray}

\begin{figure}[htbp]
\begin{center}
\noindent\epsfig{file=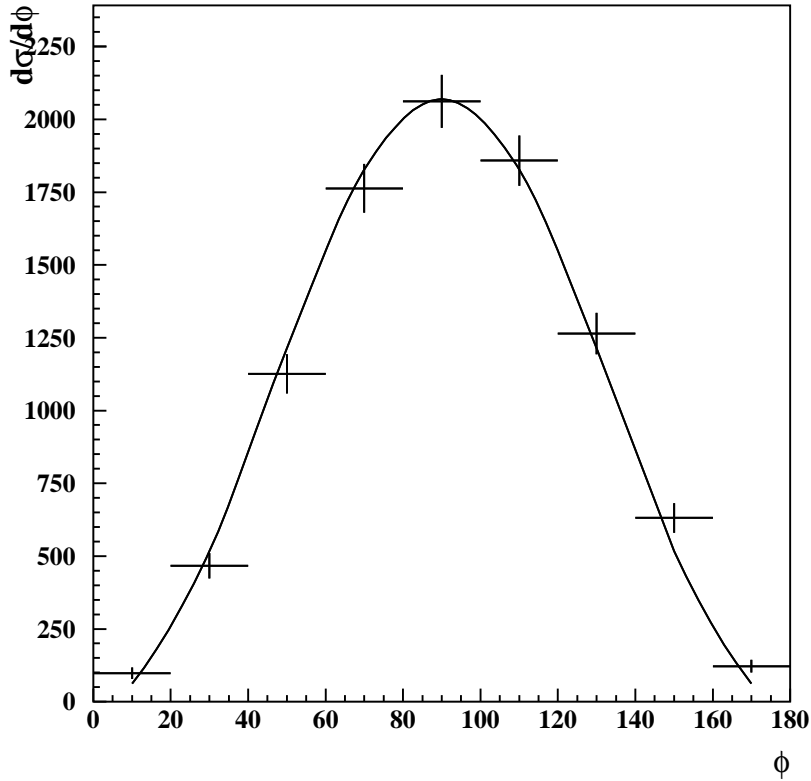,width=0.80\textwidth}
\caption{The $\sin^2 \phi $ prediction compared to $\eta'$ production
at WA102 \cite{WA}.
\label{fig1}
}
\end{center}
\end{figure}
\noindent
{\bf (i)} $\bJ^\bP \beq \bzero^\bminus$\hfill\\
The pseudoscalar cross sections behaves as
$ \d\sigma \sim  t_1\, t_2\, A_{++}^2\, \sin^2 \phi$. 
The $\phi$ dependence is hence independent of the $0^-$ being 
a glueball or a quarkonium state \cite{Castoldi,Kochelev}
and thereby provides an immediate
test of the Pomeron transformation properties.
The observed $\eta$ and $\eta'$ indeed have the $\sin^2\phi$ behaviour 
(see, for example, Figure~\ref{fig1}).
They also exhibit the $t\, e^{b\, t}$ behaviour 
\cite{WA}. 
There is the interesting possibility that glueball production 
might have a compensating $(t_1\, t_2)^{-1}$ Pomeron--Pomeron--glueball
vertex that could give a finite cross section as $t_{1,2} \to 0$ 
and provide a dynamical discrimination.

The $t_1\, t_2$ factor in (\ref{eq:zminus}) originates from the fusion 
of two transversely-polarized currents (TT) and is hence independent
of whether the current is conserved or not. 
The question of current conservation becomes testable
if longitudinal polarization (L) can contribute,
as in the case of all other states,

The $1^+$ state is interesting because Bose symmetry 
($A_{++} \propto (t_1-t_2)^2$ in (\ref{eq:oplus})) suppresses the
TT part (with helicity zero) and leaves TL 
(with helicity one)
dominant. 
As such this becomes sensitive to (non)-conserved vector effects. 

In order to exploit current (non)-conservation we consider three 
scenarios:\hfill\\[1ex]
\noindent
{\it i) Current conservation} (Model~C):
\begin{equation}
  q \cdot M =  0 
\ .
\label{eq:modelC}
\end{equation}
\noindent
{\it ii) Non-conservation} (Model~B):
\begin{equation}
  q \cdot M =  O(1) 
\ ,
\label{eq:modelB}
\end{equation}
\noindent
{\it or} (Model~A):
\begin{equation}
  q \cdot M =  O(\sqrt{-t})
\ .
\label{eq:modelA}
\end{equation}
We shall show how data discriminate among these alternatives 
and provide a consistent solution. In (\ref{eq:modelC}--\ref{eq:modelA})
$q$ ($t = q^2$) denotes the four-momentum of the current
(i.e.\ $q=q_1$ or $q_2$). Since the longitudinal polarization
vector $\epsilon_{\L} \sim q/\sqrt{-t}$ as $t \to 0$ we obtain 
$\epsilon_{\L}\cdot M \sim q\cdot M /\sqrt{-t}$ 
in contrast to $\epsilon_{\pm}\cdot M = O(1)$ 
at small $|t|$. Model C thus corresponds to the conserved-vector 
hypothesis (CVC): $M_{\L} / M_{\T} \sim \sqrt{-t}/ \mu$, where
$\mu$ is a mass scale. 
Model B is what was argued \cite{Diehl} to correspond
to the soft (``Donnachie--Landshoff'') Pomeron, 
$M_{\L} / M_{\T} \sim \mu/ \sqrt{-t}$ 
where $\mu \simeq M$, the mass of the produced meson. 
Contrary to \cite{Diehl}, we anticipate that $\mu$ is a rather small
mass scale, related to constituent binding or to instanton 
size.
Finally, model~A is a possible further alternative 
where longitudinal and transverse amplitudes have similar strengths,
$M_{\L} / M_{\T} \sim 1$. 
We now illustrate this in the case of $1^{++}$.

\noindent
{\bf (ii)} $\bJ^\bP \beq \bone^\bplus$\hfill\\
Here we concentrate on the fusion of two identical currents 
(i.e.\ photon-photon or Pomeron-Pomeron). 
Then $\sigma_0$ is as for the pseudoscalars but obeying Bose symmetry
\begin{equation}
 t_1\, t_2\, \sigma_0 =  t_1\, t_2\, \mu^2\, \hat{A}_{++}^2\, 
  \frac{ (t_1 - t_2)^2 }{\mu^4}\, \sin^2 \phi
\ .
\label{eq:onepluszero}
\end{equation}
Here and in the following the hatted quantities denote the residual,
dimensionless amplitudes, which, in general, 
are of order one\footnote{That is we generally assume that
$\hat{A}_{\lambda \lambda'} \approx \pm 1$ at small $|t_i|$. 
Additional dynamics might change this and yield, for example, 
$\hat{A}_{\L\L} = c\, \kT/\mu$.}.
As for the pseudoscalars, 
the helicity-zero part is independent of the model assumption. 

Bose symmetry allows us to determine the sign of $\xi_2$
in $\sigma_1$, (\ref{eq:approxsig},\ref{eq:oplus}), and we find
\begin{eqnarray}
  t_1\, t_2\, \sigma_1 = 
 & \mu^2\, t_1\, t_2\, 4\, \sin^2 \frac{\phi}{2}\, \hat{A}_{+L}^2
 & \mathrm{model}\,\mathrm{A}
\nonumber\\
 & \mu^4\, \kT^2\,  \hat{A}_{+L}^2 
 & \mathrm{model}\,\mathrm{B}
\nonumber\\
 & t_1\, t_2\, \kT^2\, 
            \hat{A}_{+L}^2
 & \mathrm{model}\,\mathrm{C}
\ .
\label{eq:oneplusone}
\end{eqnarray}

Recall that $\kT \to 0$ implies $\phi \to 0$ but $\phi \to 0$ yields 
$\kT \to 0$ only if $t_2/t_1 \to 1$. We can make this manifest by 
introducing the following quantities:
\begin{equation}
  \epsilon  =  \frac{\sqrt{t_1\, t_2}}{\mu^2}
\ , \qquad 
  r  =  \sqrt{ \frac{ t_2 }{ t_1 } }
\label{eq:rdef}
\end{equation}
so that
\begin{equation}
  \frac{ t_1 - t_2 }{\mu^2 }  =  \epsilon\, \frac{1 - r^2}{r}
\ ,
\qquad
  \frac{ \kT^2 }{\mu^2}  =  4\, \epsilon\, 
  \left( \sin^2 \frac{\phi}{2} + \frac{ (1-r)^2 }{4\, r} \right)
\ .
\label{eq:kTrewrite}
\end{equation}
This allows us to write the $1^+$ cross section as
\begin{equation}
 \sigma \sim \epsilon\, \left( 2\, \mu^3\, \hat{A}_{+\L} \right)^2\, 
  p_1\, \left\{ \sin^2 \frac{\phi}{2} + p_2 + p_3\, \sin^2 \phi \right\} 
\ ,
\label{eq:onepp}
\end{equation}
where
\begin{center}
\begin{minipage}[t]{0.4\textwidth}
\begin{tabular}{c|ccc}
  & \multicolumn{3}{c}{Variable}
\\
  Model & $p_1$ & $p_2$ & $p_3$
\\ \hline
 A  & $\epsilon$   & $0$      & $\epsilon^2\, F$
\\
 B  & $1         $ & $\kappa$ & $\epsilon^3\, F$
\\
 C  & $\epsilon^2$ & $\kappa$ & $\epsilon\, F$
\end{tabular}
\end{minipage}
\begin{minipage}{0.5\textwidth}
$$ \kappa =  \frac{(1 - r)^2}{4\, r}$$
\begin{equation} 
F = \frac{1}{4}\, \left( \frac{1 - r^2}{r} \right)^2\, 
  \left( \frac{ \hat{A}_{++} }{ \hat{A}_{+\L} } \right)^2
\ .
\label{eq:Fdef}
\end{equation}
\end{minipage}
\end{center}
Note that $p_3$ characterizes the strength of the helicity-zero 
part relative to the helicity-one component.

The kinematics of the WA102 experiment is such that in average 
$r \geq 0.6 $, i.e.\ close to unity. 
We make the following observations.
\begin{itemize}
\item Helicity-one dominance in all cases for integrated cross sections. 
This is seen in the WA102 data 
\cite{WA}.
It is amusing that a dominant helicity-one component
was observed already at the ISR 15 years ago \cite{ISR}. 
This was regarded as an ``unusual feature'' and, to the best of our
knowledge, has remained a puzzle until now. 
We predict an enhanced helicity-zero part 
for $\phi$ around $\pi/2$ and asymmetric $t$ values (implying small $F$). 
\item Vanishing cross sections for $\kT \to 0$ in all cases, also 
  in agreement with data \cite{WA,Close} 
  (recall that $\kT \to 0$ implies $r \to 1$). 
\item A strong $t_1\, t_2$ suppression for model A and an even stronger one
for model C, which is not observed in the data. There are, however, 
indications \cite{AK3} that data do not simply follow an 
$e^{b\, t}$ distribution but exhibit a weak turn-over at small $t$, 
precisely as is predicted in model B.
Hence we favour model~B. 
\item Owing to the smallness of the helicity-zero part ($p_3$) we expect
a dominant $\sin^2(\phi/2)$ distribution for model A. 
Indeed, since $r$ is close to unity, this is also the dominant 
behaviour for models B and C, modulated by an additional isotropic term.
\end{itemize}

\begin{figure}[htbp]
\begin{center}
\noindent\epsfig{file=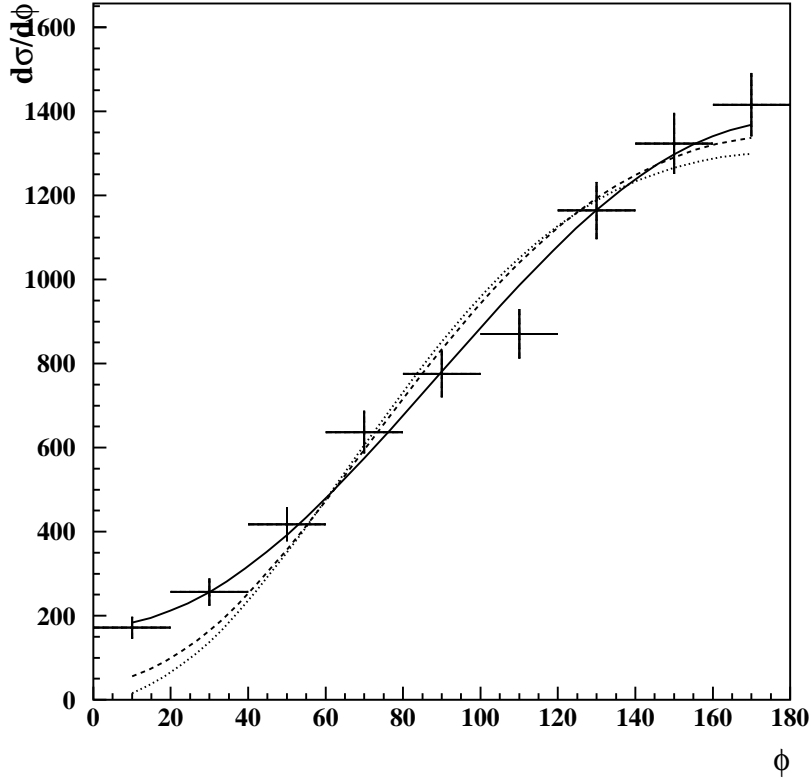,width=0.80\textwidth}
\caption{The $\phi$ distribution of $\f_1(1420)$ production at WA102\cite{WA}
fitted to 
(\ref{eq:onepp}). 
Solid line: the form of models B and C 
with $p_2$ and $p_3$ as free parameters 
(fit values are $p_2 = 0.15$ and $p_3 = 5 \times 10^{-3}$); 
Dashed line: ditto but with $p_2 = \kappa = 0.03$ fixed to the theoretical 
value (\ref{eq:Fdef}) for $r = 0.78$ (fit value is $p_3 = 0.11$);
Dotted line: the form of model A, i.e.\ $p_2 = 0$, with 
$p_3$ as free parameter (fit value is $p_3 = 0.15$).
\label{fig2}
}
\end{center}
\end{figure}
Figure~\ref{fig2} shows a comparsion with the $\f_1(1420)$ data from WA102 
exhibiting the dominance of the $\sin^2(\phi/2)$ term. An excellent
description is described in both model~B, or in~C,  
with a small isotropic term ($p_2 = 0.15$) and a very small
helicity-zero contribution ($p_3 = 0.005$).
If we fix $p_2$ to its theoretical value using 
$\langle -t_1\rangle = 0.145\,$GeV$^2$ and 
$\langle -t_2\rangle = 0.240\,$GeV$^2$, we can determine the mass scale
$\mu$ by taking $\hat{A}_{++} / \hat{A}_{+\L} = 1$. 
We find $\mu \approx 0.34\,$GeV for $\f_1(1285)$ 
and $\mu \approx 0.40\,$GeV for $\f_1(1420)$ based on the fit values
$p_3 = 0.27$ and $0.11$, respectively.

\begin{figure}[htbp]
\begin{center}
\noindent\epsfig{file=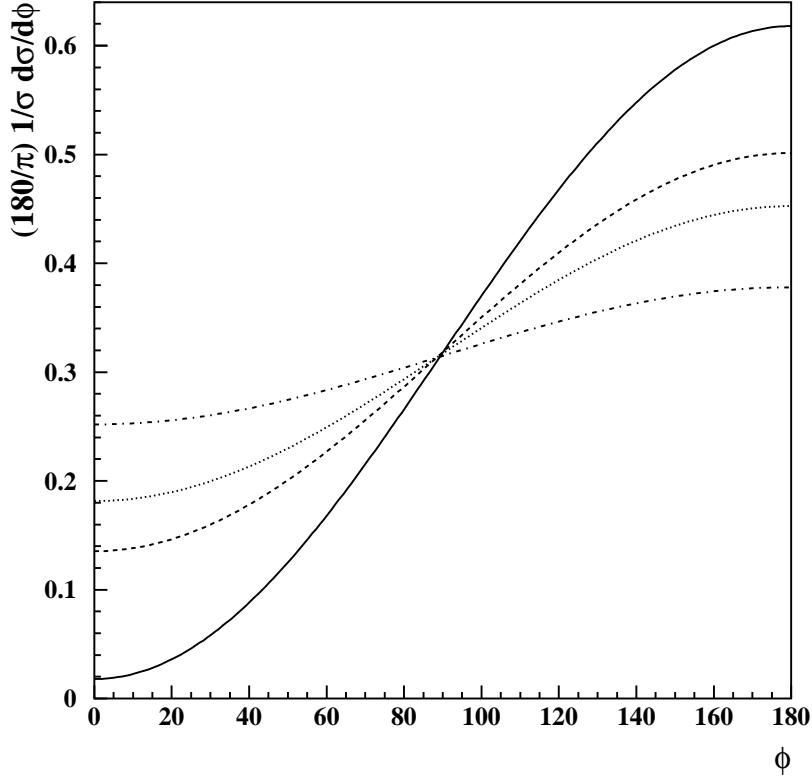,width=0.80\textwidth}
\caption{Predictions for the normalized $\phi$ distribution 
of $1^+$ production for various values of $t_2/t_1$: 
$0.5$ (solid), $0.1$ (dashed), $0.05$ (dotted), $0.01$ (dashed-dotted).
The helicity-zero contribution has been neglected.
\label{fig3}
}
\end{center}
\end{figure}
To test our prediction further,  we propose that experimental data on
the $\phi$ distributions should be analyzed for various $r$ bins. 
Lowering $r$ from $1$ to $0$ we predict a change from $\sin^2(\phi/2)$ to 
a distribution essentially flat in $\phi$, see Figure~\ref{fig3}.

We now turn to scalars and tensors; these are interesting as both
potential glueballs as well as established quarkonia are known to arise. 
Further it is known that, both in the scalar and tensor sector, there 
are some mesons that are suppressed at low $\kT$ and others that
do not show this feature \cite{WA}. Therefore 
it is interesting to separate the Pomeron from the meson dynamics. 

\noindent
{\bf (iii)} $\bJ^\bP \beq \bzero^\bplus$\hfill\\
Obviously $\sigma_2 = \sigma_1 = 0$ but 
$\sigma_0 = (A_{LL} - \xi_1\, A_{++}\,\cos\phi)^2$ depends on the model:
\begin{eqnarray}
 \d\sigma \sim 
 & \mu^2\, t_1\, t_2\, \hat{A}_{++}^2\, \left( R - \cos\phi \right)^2 
 & \mathrm{model}\,\mathrm{A}
\nonumber\\
 & \mu^6\, \hat{A}_{\L\L}^2\,
   \left( 1 - \frac{\epsilon}{R}\, \cos\phi \right)^2 
 & \mathrm{model}\,\mathrm{B}
\nonumber\\
 & \mu^2\,t_1\, t_2\, A_{++}^2\,
  \left( R\, \epsilon  - \cos\phi \right)^2 
 & \mathrm{model}\,\mathrm{C}
\ .
\label{eq:zeroplus}
\end{eqnarray}
For all cases, 
\begin{equation}
  R = \xi_1\, \frac{ \hat{A}_{\L\L} }{ \hat{A}_{++} }
\label{eq:Rdef}
\end{equation}
is, in general\footnote{See footnote~1.},
a number with absolute value of order one, $|R| = O(1)$. 
Note that the sign of $R$ cannot be fixed from first principles. 

Experimentally the $t_i$ distributions seem to continue to grow
at small $|t_i|$ \cite{WA76}
indicating that non-conserving parts must be present
in the cross section. 
In model B we have the possibility to compensate the factor $t_1\, t_2$
in (\ref{eq:dsigde}) through the $1\sqrt{-t_i}$ enhanced longitudinal
amplitude. 

\begin{figure}[htb]
\begin{center}
\noindent\epsfig{file=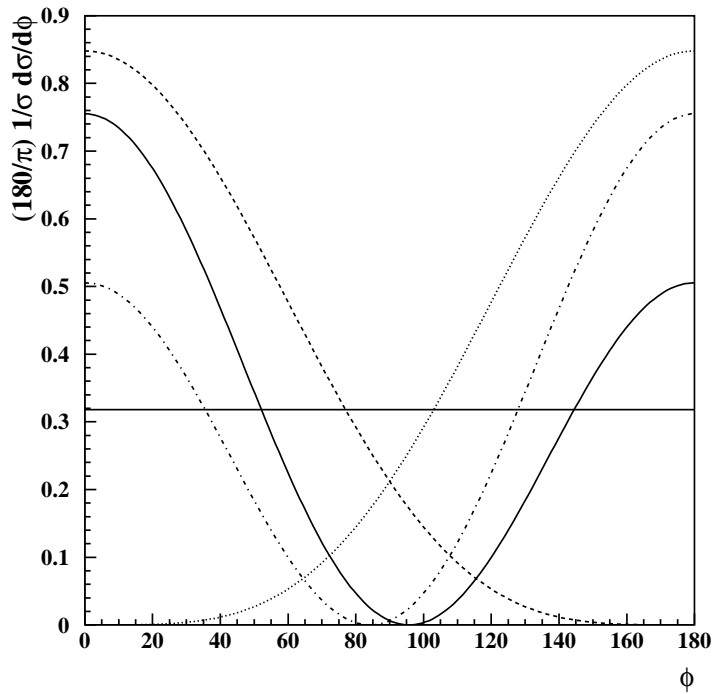,
width=0.70\textwidth}
\caption{Predictions for the normalized $\phi$ distribution 
of $0^+$ production in model~B, 
$\d\sigma/\d\phi \sim (1 - \delta\, \cos\phi)^2$, 
for various values of $\delta = \epsilon / R$: 
$-10$ (solid), $-1$ (dashed), $0$ (solid), 
$+1$ (dotted), $+10$ (dashed-dotted).
\label{fig4}
}
\end{center}
\end{figure}

Focussing now on model~B we note that the $\phi$ dependence is very sensitive
to the ratio $\delta \equiv \epsilon / R = \sqrt{t_1\, t_2}/(R\, \mu^2)$. 
The $\phi$ distribution changes from isotropic at small $\delta$ 
to $\cos^2 \phi$ at large $\delta$. Data thus
allow the determination of the size of $\mu$. The interesting regime 
is when $| \delta | \approx 1$. If, as suggested by the $1^+$ analysis, 
$\mu$ is of the order of $\Lambda_{\mathrm QCD}$, the constituent-quark
mass, the average $\kT$, or the inverse instanton size, then this would
occur for the typical $t_i \sim 0.2$ of WA102. In such a case, depending
on whether $\delta=\pm 1$, one expects $\cos^4(\phi/2)$ or $\sin^4(\phi/2)$, 
see Figure~\ref{fig4}. 

Data on $\f_0(980)$ and $\f_0(1500)$ show no suppression at small $\kT$
\cite{WA}. This implies $\delta$ is negative  
and we predict a dominant $\cos^4(\phi/2)$ dependence. 
Conversely, the $\f_0(2000)$, which is observed to vanish as
$\kT \to 0$ \cite{WA}, will have $\delta$ positive 
and hence be maximum as $\phi \to \pi$.
To further probe the dynamics of scalar production we propose
that the $\phi$ distributions of the data be analyzed for various bins of 
$\sqrt{t_1\, t_2}$, 
see Figure~\ref{fig4}. 

\noindent
{\bf (iv)} $\bJ^\bP \beq \btwo^\bplus$\hfill\\
The cross section in the various models is given by
\begin{eqnarray}
\d\sigma^{\mathrm B} & \sim &
  \mu^6\, \hat{A}_{\L\L}^2\, \left\{
   \left( 1 - \frac{\epsilon}{R}\, \cos\phi \right)^2 
   + \frac{\kT^2}{\mu^2}\,  
        \left(\frac{\hat{A}_{+\L}}{\hat{A}_{\L\L}} \right)^2 
   + \frac{1}{2}\, \epsilon^2\, 
        \left(\frac{\hat{A}_{+-}}{\hat{A}_{\L\L}} \right)^2 \right\}
\label{eq:twoB}\\
\d\sigma^{\mathrm A} & \sim &
  \mu^2\, t_1\, t_2\,  \left\{ \hat{A}_{++}^2\, 
   \left( R - \cos\phi \right)^2 
   + \hat{A}_{+\L}^2\, 4\, \sin^2 \frac{\phi}{2} 
   + \frac{1}{2}\, \hat{A}_{+-}^2 \right\}
\label{eq:twoA}\\
\d\sigma^{\mathrm C} & \sim &
  \mu^2\, t_1\, t_2\,  \left\{ \hat{A}_{++}^2\, 
   \left( \epsilon\, R - \cos\phi \right)^2 
   + \frac{\kT^2}{\mu^2}\, \hat{A}_{+\L}^2 
   + \frac{1}{2}\, \hat{A}_{+-}^2 \right\}
\ .
\label{eq:twoC}
\end{eqnarray}
The three terms for each case correspond to meson helicity zero, one,
and two, respectively. Here we have again made use of Bose symmetry to
fix the sign in $\sigma_1$. 

In CVC, the residual amplitude $\hat{A}_{+-}$ is naturally of order one;
hence a large helicity-two component is expected for quarkonium states. 
The central-production data \cite{WA} do not agree with this even though 
helicity-two dominance is well established and understood
for $\e^+\e^-$ \cite{ee,hel2}. 
This marked difference was commented on in our earlier
paper and is another motivation for non-CVC dynamics. 

As for scalar-meson production, experimental hints 
\cite{WA76}
for a continuous 
growth of the $t$ distributions at small $|t|$ favour the
current non-conserving alternative B. Let us see whether we can find
a consistent picture.

Well-established quarkonia are known to be suppressed at small $\kT$
\cite{WA}. Referring to (\ref{eq:twoB}) this implies that 
\begin{equation}
\epsilon^2\, (\hat{A}_{+-}/\hat{A}_{\L\L})^2 \ll 1 \qquad
\mathrm{and} \quad 
\delta = \epsilon/ R \approx + 1
\ .
\label{eq:condi}
\end{equation}
If this was the case
we would predict a helicity  hierarchy, namely 
$\sigma_0 \gg \sigma_1 \gg \sigma_2$. This looks consistent since
preliminary WA102 data on $\f_2(1270)$ and 
$\f_2'(1525)$ \cite{AK6} seem to support this 
ordering. 
Moreover, the condition $| \delta | \approx 1$ is also what we 
already found from our analysis of scalars. 

If $\delta \approx +1$ (in accord with the suppression as $\kT \to 0$) 
then it is sensible to expand in powers
of $\sin^2(\phi/2)$
\begin{equation}
 \d\sigma^{\mathrm{B}} \sim 4\, \mu^6\, \hat{A}_{\L\L}^2\, \left\{
  p_2\, \sin^4\frac{\phi}{2} +   p_1\, \sin^2\frac{\phi}{2} +   p_0 \right\}
\ ,
\end{equation}
where
\begin{eqnarray}
 p_2 & = & \delta^2
\nonumber\\
 p_1 & = & (1-\delta) + \epsilon\, 
  \left( \frac{\hat{A}_{+\L}}{\hat{A}_{\L\L}} \right)^2
\nonumber\\
 p_0 & = & (1-\delta)^2 + \frac{\epsilon}{4}\, \frac{(1-r)^2}{r}\, 
  \left( \frac{\hat{A}_{+\L}}{\hat{A}_{\L\L}} \right)^2
  + \frac{\epsilon^2}{8}\, 
  \left( \frac{\hat{A}_{+\-}}{\hat{A}_{\L\L}} \right)^2
\ .
\end{eqnarray}
We see that helicity zero contributes to all $p_i$, 
helicity one to $p_1$ and $p_0$, and helicity two to $p_0$ only.
Hence we predict that both the helicity structure and the
$\phi$ dependence should vary with $t$.

If, as has been suggested elsewhere \cite{CK}, $2^+$ glueballs survive 
as $\kT \to 0$, then we would expect a $\phi$ distribution more similar 
to $\f_0(980, 1500)$ than to $\f_2(1270)$. 

\noindent
{\bf (v)} $\bJ^\bP \beq \btwo^\bminus$\hfill\\
The cross section in model~B is given by
\begin{equation}
 \d\sigma[2^-] \sim  t_1\, t_2\, \mu^2 \hat{A}_{++}^2\, \sin^2 \phi
  + \mu^4\, \pT^2 \hat{A}_{+\L}^2 + 
 \frac{t_1\, t_2\, (t_1 - t_2)^2 }{2\,  \mu^2} \hat{A}_{+-}^2
\ ,
\label{eq:twominus}
\end{equation}
where the three terms on the rhs correspond to helicity zero, one, and two, 
respectively. The expression for models A and C are given by replacing
the helicity-one term by (\ref{eq:oneplusone}) with the substitutions 
$\kT \mapsto \pT$ and $\sin(\phi/2) \mapsto \cos(\phi/2)$.
We can write (\ref{eq:twominus}) as
\begin{eqnarray}
\d\sigma[2^-] & \sim & \epsilon\, \mu^6\, \left\{
  \epsilon\, \hat{A}_{++}^2\, \sin^2\phi 
 + 4\, \hat{A}_{+\L}^2 \left[ \cos^2 \frac{\phi}{2} + \frac{(1-r)^2}{4\,r}
  \right]
\right. \nonumber\\ & &~ \left. \qquad +~ 
 \frac{\epsilon^2}{2}\, \left( \frac{ 1 - r^2}{r} \right)^2\,
   \hat{A}_{+-}^2
 \right\}
\ .
\end{eqnarray}
Experimentally the $2^-$ states are known to be suppressed at 
small $\kT$ \cite{WA}. This implies that the reduced helicity-one amplitude 
$\hat{A}_{+\L}(\kT \to 0)$ is suppressed. 
This happens in the non-relativistic quark model 
coupling to two photons, where $\hat{A}_{+\L}$ is identically zero.
If this is also true in the QCD case, 
then we predict:
\begin{itemize}
\item Cross sections that behave as $t\, e^{b\, t}$ at small $t$, 
similar to the $0^-$ ones.
\item A small helicity-one contribution. 
\item A helicity-zero contribution that behaves as $\sin^2\phi$.
\item A helicity-two contribution that is isotropic in $\phi$. 
\item A ratio $\sigma_2/\sigma_0$ that vanishes for $t_2 = t_1$ and 
increases with increasing difference $|t_2 - t_1|$. 
\end{itemize}

In summary, we eagerly await new results on the $\phi$ and $t$ dependences 
for central meson production enabling the parameters, as described 
in this paper, to be determined. 
Once these parameters are determined, the dynamical nature of the mesons 
will become clear. 
\hfill\\[0.1ex]

\noindent
{\it Acknowledgements:}\\[0.5ex]
We thank A.\ Donnachie, A.\ Kaidalov, A.\ Kirk, N.\ Kochelev 
and P.\ Landshoff for discussions.


\begin{thebibliography}{99}

\bibitem{CK}
 F.E.\ Close and A.\ Kirk, \PLB{397}{1997}{333}.

\bibitem{Robson}
 D.\ Robson, \NPB{130}{1977}{328};
 \hfill\\
 F.E.\ Close, \RPP{51}{1988}{833}.

\bibitem{WA}
 WA102 Collaboration (A.\ Kirk et al.), \hep{9810221};\hfill\\
 WA102 Collaboration (D.\ Barberis et al.),  \PLB{440}{1998}{225},
     \PLB{432}{1998}{436}, \PLB{427}{1998}{398}, \PLB{397}{1997}{339}.

\bibitem{Us}
 F.E.\ Close and G.A.\ Schuler, 
  \hep{9902243}, Physics Letters B in press.

\bibitem{Diehl}
T.\ Arens, O.\ Nachtmann, M.\ Diehl, P.V.\ Landshoff,
 \ZPC{74}{1997}{651}.

\bibitem{kaid}
A.B.\ Kaidalov (private communications).

\bibitem{Castoldi}
 P.\ Castoldi, R.\ Escribano and J.M.\ Fr\`{e}re, \PLB{425}{1998}{359}.

\bibitem{Kochelev}
N.I.\ Kochelev,  \hep{9902203};\hfill\\
N.I.\ Kochelev, T.\ Morii and A.V.\ Vinnikov, \hep{9903279}. 

\bibitem{ISR}
P.\ Chauvat et al., \PLB{148}{1984}{382}. 

\bibitem{Close}
 F.E.\ Close, \PLB{419}{1998}{387}.

\bibitem{AK3}
A.\ Kirk (private communications).

\bibitem{WA76}
WA76 Collaboration (T.A.\ Armstrong et al.), \ZPC{51}{1991}{351}. 

\bibitem{ee}
See, for example, M.\ Poppe, 
Int.\ J.\ Mod.\ Phys.\ {\bf A1} (1986) 612. 

\bibitem{hel2}
F.E.\ Close, Nucl.\ Phys.\ {\bf B} (proc.\ suppl.) {\bf 21} (1991) 423;\hfill\\
T.\ Barns, F.E.\ Close and Z.\ Li, \PRD{43}{1991}{2161};\hfill\\
G.A.\ Schuler, F.A.\ Berends and R. van Gulik, \NPB{523}{1998}{423}.

\bibitem{AK6}
WA102 Collaboration (D.\ Barberis et al.), 
\hex{9903042}, \hex{9903043}, to be published in Phys.\ Lett.\ B.

\end{thebibliography}
\end{document}